\documentclass[rnote]{aa}  
\usepackage{graphicx}
\usepackage{natbib}
\bibpunct{(}{)}{;}{a}{}{,}
\usepackage{txfonts}
\newcounter{Rco}
\newcommand{\Ionst}[1]{\setcounter{Rco}{#1}\Roman{Rco}}
\newcommand{\Ion}[2]{\mbox{#1\ {\scriptsize\Ionst{#2}}}}

\newcommand{\loggw}[1]{\mbox{$\log g\hspace{-0.5mm} =\hspace{-0.5mm}  #1$}}

\newcommand{\ab}[1]{\mbox{Fig.\,\ref{#1}}}
\newcommand{\sA}[1]{\mbox{(Fig.\,\ref{#1})}}

\newcommand{\gla}{\raisebox{-0.10em}{$\stackrel{>}{{\mbox{\tiny $\sim$}}}$}}

\newcommand{\Teff}{\mbox{$T_\mathrm{eff}$}}
\newcommand{\Teffw}[1]{\mbox{$\Teff\hspace{-0.5mm}=\hspace{-0.5mm}#1\,\mathrm{K}$}}
\newcommand{\hz}{\object{HZ\,43}}
\newcommand{\lsv}{\object{LS\,V\,$+46\degr 21$}}
\newcommand{\pnsh}{\object{Sh\,2$-$216}}
\newcommand{\sir}{\object{Sirius\,B}}
\renewcommand{\object}{{}}
\begin{document}
\title{Uncertainties in (E)UV model atmosphere fluxes}
\author{T\@. Rauch}

\institute{Institute for Astronomy and Astrophysics,
           Kepler Center for Astro and Particle Physics,
           Eberhard Karls University, 
           Sand 1,
           D-72076 T\"ubingen, 
           Germany,
           \email{rauch@astro.uni-tuebingen.de}}


\date{Received 21 January 2008 / Accepted 4 February 2008}


\abstract
         {
          During the comparison of synthetic spectra calculated with 
          two NLTE model atmosphere codes, namely \emph{TMAP} and \emph{TLUSTY},
          we encounter systematic differences in the EUV fluxes due to the
          treatment of level dissolution by pressure ionization.
         }
         {
          In the case of \sir, we demonstrate an uncertainty in modeling the
          EUV flux reliably in order to
          challenge theoreticians to improve the theory of
          level dissolution.
         }
         {We calculated synthetic spectra for hot, compact stars
          using state-of-the-art NLTE model-atmosphere techniques.}
         {
          Systematic differences may occur due to a code-specific cutoff frequency of
          the \Ion{H}{1} Lyman bound-free opacity. This is the case for \emph{TMAP} and \emph{TLUSTY}.
          Both codes predict the same flux level at wavelengths lower than about 1500\,\AA\
          for stars with effective temperatures (\Teff) below about 30\,000\,K only, if the 
          same cutoff frequency is chosen.
         }
         {
          The theory of level dissolution in high-density plasmas, which is
          available for hydrogen only should be generalized to all species.
          Especially, the cutoff frequencies for the bound-free opacities should
          be defined in order to make predictions of UV fluxes more reliable.
         } 


\maketitle

\section{Introduction}
\label{sect:introduction}

NLTE model atmosphere codes for hot, compact stars have arrived at a high level of
sophistication and are successfully employed for spectral analyses, 
e.g\@. the T\"ubingen NLTE Model-Atmosphere Package 
\emph{TMAP}\footnote{http://astro.uni-tuebingen.de/\raisebox{.2em}{{\tiny $\sim$}}rauch/TMAP/TMAP.html}
\citep{wea2003, rd2003} in the case of \lsv, the central star of \pnsh\ \citep{rea2007}.

In the case of high-gravity stars like \lsv\ (\loggw{6.9} [cm/sec$^2$], \Teffw{95\,000}) or
neutron stars \citep[e.g.][]{sw2007}, the consideration of the 
dissolution of atomic levels by plasma perturbation \citep[for details see, e.g., ][]{HM88} 
is crucial for a reliable model atmosphere calculation \citep{HHL1994}. 

\citet{bbr2006, bbr2008} established the DA-type white dwarfs 
\hz\  (\loggw{7.9}, \Teffw{51\,111}) and 
\sir\ (\loggw{8.6}, \Teffw{24\,897}) 
as soft X-ray standards. 
For a cross-calibration between the Chandra LETG + HRC-S, the EUVE spectrometer and the ROSAT PSPC,
pure hydrogen \emph{TMAP} model atmospheres and synthetic spectra were used. In the case of \sir\ with a 
higher surface gravity $g$ (\loggw{8.6}), the level dissolution due to pressure ionization is even 
more efficient \sA{fig:leveldiss} and thus, it is more important to consider it properly.

\begin{figure}[ht]
  \resizebox{\hsize}{!}{\includegraphics{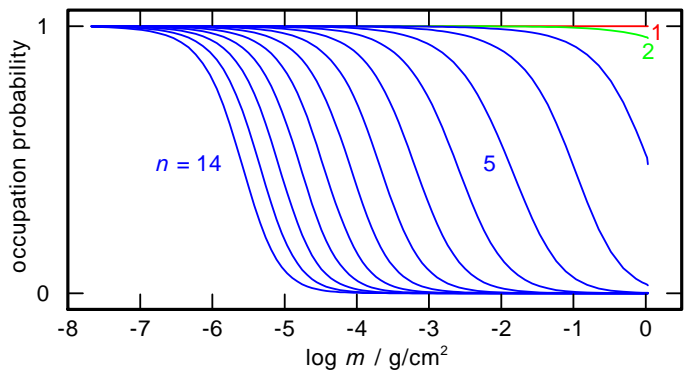}}
  \caption[]{Occupation probabilities of the lowest 14 \Ion{H}{1} levels considered in our \emph{TMAP} 
             model atmosphere calculation for \sir. Note that the line
             formation takes place at $\log m \approx -5.5 - -3.0$.}
  \label{fig:leveldiss}
\end{figure}

\section{Level dissolution}
\label{sect:leveldiss}

Recently, Jelle Kaastra (priv\@. comm.) has drawn our attention to a deviation between
\emph{TMAP} and 
\emph{TLUSTY}\footnote{http://nova.astro.umd.edu} 
fluxes for \sir\ \sA{fig:discrep} at wavelengths lower than about 1\,500\AA,
while in the case of \hz, the model fluxes are in agreement.

\begin{figure}[ht]
  \resizebox{\hsize}{!}{\includegraphics{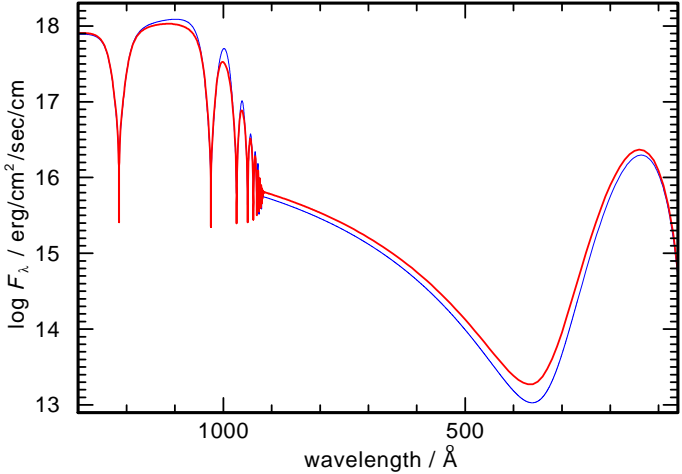}}
  \caption[]{Discrepancy between synthetic spectra for \sir\ calculated by
             \emph{TMAP}   (thick, long cutoff, see text) and 
             \emph{TLUSTY} (thin, short cutoff, Lanz priv\@. comm.) with the same parameters.}
  \label{fig:discrep}
\end{figure}

In the last two decades, we have thoroughly compared \emph{TMAP} and \emph{TLUSTY} from time to time 
and found only negligible differences due to different numerical approaches and slightly
different constants. 
In an investigation of the flux discrepancy, we are now able to identify
its reason. Both codes, \emph{TMAP} as well as \emph{TLUSTY}, follow a generalized form 
to consider the level dissolution by \citet{HHL1994}. A hitherto unsolved problem, however, 
is a precise, generalized formulation of the extrapolation of the hydrogen Lyman bound-free opacity 
into a pseudo-continuum below the unperturbed position of the absorption threshold \citep[cf\@.][]{dam87}. 
\emph{TMAP} uses a heuristic approach with

\begin{equation}
\nu^\mathrm{th}_\mathrm{i} = f \cdot
   \left[\frac{1}{n_\mathrm{i}^2} - \frac{1}{(n_\mathrm{i}+1)^2}\right] \cdot \nu_\mathrm{i}^\mathrm{th,0}
\end{equation}

\noindent
where $\nu^\mathrm{th}_\mathrm{i}$ and $\nu^\mathrm{th,0}_\mathrm{i}$ are the 
extrapolated and unperturbed threshold frequencies, respectively, of level $i$ and 
$n_\mathrm{i}$ is its principal quantum number, i.e., $n_\mathrm{i}^2=1$ for
the Lyman continuum. Since this approach results in artificial absorption edges at 
$f = 1$   (corresponding to $\lambda^\mathrm{th}_1 = 1\,215.67.\,\mathrm{\AA}$,
``short'' cutoff), 
\emph{TMAP} introduced 
$f = 0.5$ (corresponding to $\lambda^\mathrm{th}_1 = 2\,431.34\,\mathrm{\AA}$,
``long'' cutoff)
in order to achieve a smooth transition into the continuum \sA{fig:artedges}.

\begin{figure}[ht]
  \resizebox{\hsize}{!}{\includegraphics{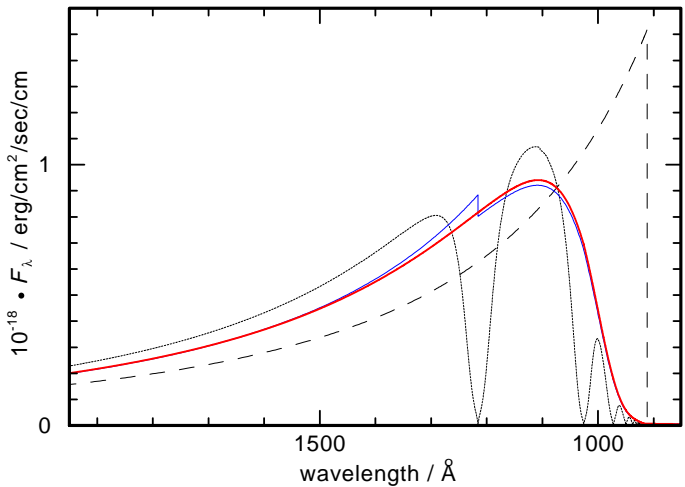}}
  \caption[]{Synthetic fluxes for \sir\ calculated from continuum models (no line
             transitions considered for test reasons) with a short (thin line) and
             a long cutoff (thick).
             The dashed line shows a synthetic spectrum
             with neglection of the level dissolution.
             Note that the artificial absorption edge at the short cutoff is
             hidden in the strong L\,$\alpha$ absorption when those lines are
             considered (dotted).}
  \label{fig:artedges}
\end{figure}

\emph{TLUSTY} uses a different treatment of the continuum. Based on the physical
picture used to derive the pseudo-continuum formulation \citep[cf\@.][]{dam87}, 
which is valid only near the ionization limit, an artificial ``short'' cutoff of 
the bound-free opacity of the hydrogen Lyman-continuum at 
$\lambda^\mathrm{th}_1 = 925\,\mathrm{\AA}$ (Lanz priv\@. comm.) may be chosen (\emph{TMAP}: 
$\lambda^\mathrm{th}_1 = 2\,431.34\,\mathrm{\AA}$). A test calculation has shown that the 
model atmosphere fluxes of \emph{TMAP} and \emph{TLUSTY} agree within 5\,\% if a long 
cutoff is used by \emph{TLUSTY}, too (Lanz priv\@. comm., \ab{fig:longcutoff}).

\begin{figure}[ht]
  \resizebox{\hsize}{!}{\includegraphics{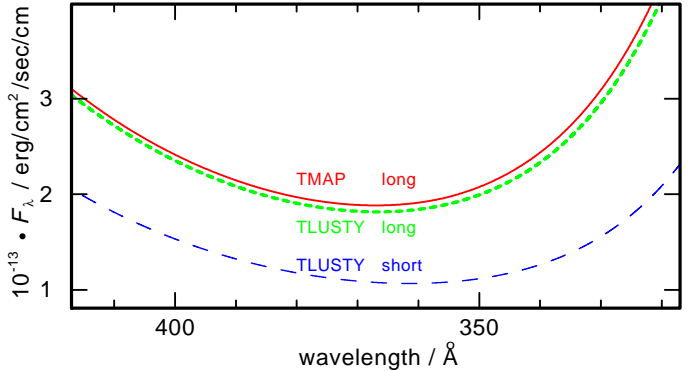}}
  \caption[]{Comparison of \emph{TMAP} and \emph{TLUSTY} 
             (dotted: long cutoff, dashed: short cutoff) fluxes in the vicinity of
             the largest deviation \sA{fig:discrep}.
            }
  \label{fig:longcutoff}
\end{figure}

Test calculations of pure hydrogen models with \emph{TMAP} via the WWW interface
\emph{TMAW}\footnote{http://astro.uni-tuebingen.de/\raisebox{.2em}{{\tiny $\sim$}}rauch/TMAW/TMAW.html}
at \loggw{8.6} and \Teff\ between
10\,000 and 50\,000\,K have shown that deviations between short-cutoff and long-cutoff
model fluxes are negligible at \Teff\,\gla\,30\,000\,K due to the increasing degree of
ionization. This explains the good agreement of \emph{TMAP} and \emph{TLUSTY} fluxes in case
of \hz.

\section{Conclusions}
\label{sect:coclusions}
Since no reliable theory is available, 
the choice of different cutoff frequencies of the \Ion{H}{1} Lyman bound-free opacity
in the NLTE model-atmosphere codes \emph{TMAP} and \emph{TLUSTY} demonstrates that
the estimate of the pseudo-continuum at longer wavelengths is presently an uncertainty 
and definitely deserves further investigation. 
However, the necessity of a cutoff in order to avoid an unrealistic opacity in the infrared
is shown in \ab{fig:fluxratio}.

\begin{figure}[ht]
  \resizebox{\hsize}{!}{\includegraphics{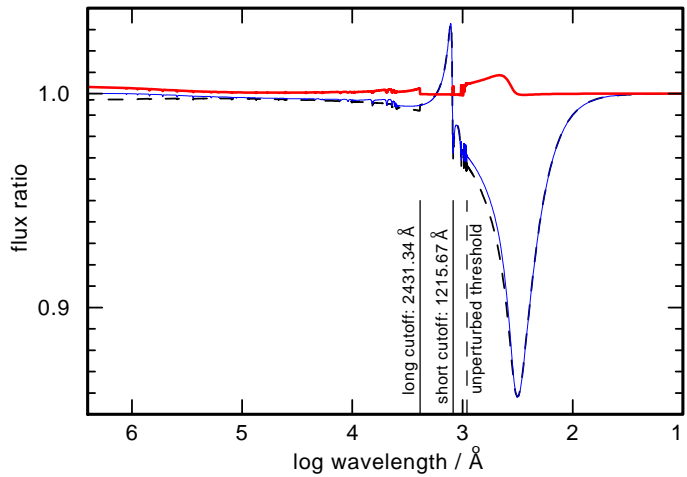}}
  \caption[]{Ratios of synthetic fluxes calculated by \emph{TMAP}. Shown are 
             short/long cutoff (dashed),
             short/extralong cutoff (thin, the extralong cutoff frequency is 10$^{12}$\,Hz), and
              long/extralong cutoff (thick).
             Note that in the case of the extralong cutoff, the flux ratio slightly increases above
             100\,000\,\AA\ due to an artificial bound-free opacity in the infrared.}
  \label{fig:fluxratio}
\end{figure}

Moreover, a reliable theory for level dissolution by pressure ionization is presently available only 
for \ion{H}{i}. Since this is important for all other species as well, 
a generalized theory is highly desirable.
However, this is out of the scope of this work.

\begin{acknowledgements}
We are indebted to Jelle Kaastra, Thierry Lanz and Ivan Hubeny who originally found 
discrepancies in the X-ray/EUV calibrations of \citet{bbr2006} between
\emph{TLUSTY} and \emph{TMAP} model-atmosphere fluxes. They informed Klaus Beuermann
about this issue ahead of their publication. This initiated a re-investigation of
our flux calibration \citep{bbr2006}. We found two trivial errors \citep{bbr2008} and
a strong systematic difference in the case of \sir. Special thanks go to Thierry Lanz
who then calculated two \emph{TLUSTY} models with different cutoff frequencies of the \Ion{H}{1} 
Lyman bound-free opacity and thus worked out the basic reason for the differences between
\emph{TLUSTY} and \emph{TMAP}.

We thank Klaus Beuermann, Klaus Werner, and Valery Suleimanov for comments and discussions.

T.R\@. is supported by the \emph{German Astrophysical Virtual Observatory} project
of the German Federal Ministry of Education and Research (BMBF) under grant 05\,AC6VTB. 
\end{acknowledgements}

\bibliographystyle{aa}
\bibliography{hm.bbl}

\end{document}